\documentclass{PoS}
\usepackage{amsmath}
\usepackage{epsfig}
\usepackage{dsfont}
\usepackage{psfrag}

\newcommand{\cO}{{\cal O}}
\newcommand{\cQ}{{\cal Q}}
\newcommand{\cR}{{\cal R}}
\newcommand{\cS}{{\cal S}}
\newcommand{\cZ}{{\cal Z}}
\newcommand{\dZ}{{\mathds{Z}}}

\newcommand{\Dslash}{\relax{\kern+.25em / \kern-.70em D}}
\newcommand{\fla}{s}
\newcommand{\vx}{\mathbf{x}}
\newcommand{\vy}{\mathbf{y}}

\PoS{PoS(LAT2005)214}

\title{$B^0-\bar B^0$ mixing in the static approximation from the
  Schr\"odinger Functional and twisted mass QCD\thanks{Preprint:
  CERN-PH-TH/2005-159, DESY 05-156}}

\ShortTitle{$B^0-\bar B^0$ mixing in the static approximation from the
  SF and tmQCD}

\author{\epsfig{file=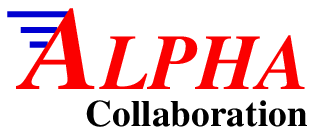, width=0.15\textwidth}}

\author{\speaker{Filippo Palombi}\\ 
        
        DESY, Theory Group, Notkestra\ss e 85, D-22603 Hamburg, Germany\\
        E-mail: \email{filippo.palombi@desy.de}}

\author{Mauro Papinutto\\
        John von Neumann-Institut f\"ur Computing NIC, Platanenallee 6, D-15738 Zeuthen, Germany\\
        E-mail: \email{mauro.papinutto@desy.de}}

\author{Carlos Pena\\
	CERN, Physics Department, Theory Division, CH-1211 Geneva 23, Switzerland\\
        E-mail: \email{carlos.pena.ruano@cern.ch}}

\author{Hartmut Wittig\\
        DESY, Theory Group, Notkestra\ss e 85, D-22603 Hamburg, Germany\\
        E-mail: \email{hartmut.wittig@desy.de}}

\abstract{We discuss the renormalisation properties of parity-odd
  $\Delta B=2$ operators with the heavy quark treated in the static
  approximation. Via twisted mass QCD (tmQCD), these operators provide
  the matrix elements relevant for the $B^0-\bar B^0$ mixing
  amplitude. The layout of a non-perturbative renormalisation
  programme for the operator basis, using Schr\"odinger Functional
  techniques, is described. Finally, we report our results for a one-loop
  perturbative study of various renormalisation schemes with
  Wilson-type lattice regularisations, which allows, in particular, to 
  compute the NLO anomalous dimensions of the operators in the SF
  schemes of interest.}

\FullConference{XXIIIrd International Symposium on Lattice Field
  Theory\\
  25-30 July 2005\\
  Trinity College, Dublin, Ireland}

\begin{document}
  
\section{Introduction}

The oscillations of the system $B^0-\bar B^0$ are one of the crucial topics
in particle physics. Their understanding represents a challenging
bridge towards the numerical determination of the Cabibbo-Kobayashi-Maskawa
(CKM) matrix and a severe test of the Standard Model. The transition
amplitude responsible for the mixing, 
\begin{align}
& \langle \bar B^0|\cO_{\rm VV+AA}|B^0\rangle =
\frac{8}{3}f_B^2m_B^2B_B,
\end{align}
is mediated by the four-quark operator $\cO_{\rm VV+AA}=(\bar
b\gamma_\mu d)(\bar b\gamma_\mu d) + (\bar b\gamma_\mu \gamma_5 d)(\bar
b\gamma_\mu \gamma_5 d)$. It has been shown that 
the renormalisation of such operators is non-trivial in Wilson-like
regularisations, resulting in a mixing with other four-quark
operators \cite{Donini:1999sf}. 
Here we propose a strategy to compute the matrix element, based on
the static approximation of the heavy quark plus the adoption of a
tmQCD regularisation for the light one. It will be proved that,
following these assumptions, the mixing under renormalisation is eliminated. 
Of course, the potential results of the proposed approach constitute
an intermediate step to the physical solution, as they must be
considered in view of the calculation of heavy quark subleading
corrections and/or interpolations  to relativistic calculations
performed at accessible heavy quark masses \cite{Be\'cirevi\'c:2001xt}.

\section{Operator mapping in tmQCD}

In order to implement our strategy, we start by fixing the
notation. The $b$ quark is replaced by an infinitely massive quark,
described by a pair of static fields $(\psi_h,\psi_{\bar h})$
propagating forward and backward in time, whose dynamics is governed by the
Eichten-Hill action \cite{Eichten:1989zv} (or one of its ALPHA
variants \cite{DellaMorte:2005yc}),  
\begin{gather}
S^{\rm stat}[\psi_h,\psi_{\bar h}] = a^4\sum_x \left[\bar\psi_h(x)\nabla_0^*\psi_h(x) -
  \bar\psi_{\bar h}(x)\nabla_0\psi_{\bar h}(x)\right].
\end{gather}
On the light quark side, the degrees of freedom are represented by an
isospin doublet $\psi_\ell^T = (u,d)$, made of an {\it up} and a {\it
  down} quark, and described according to the tmQCD action\footnote{We
  will always work in the so-called twisted basis. For a discussion of
  the problem in the physical basis, see \cite{DellaMorte:2004wn}.},  
\begin{gather}
S^{\rm tmQCD} = a^4\sum_{x}\
      \left\{\psi_{\ell}(x)\left[\Dslash + m_{\ell} +
	i\mu_{\ell}\tau^3\gamma_5\right]\psi_{\ell}(x)\right\}.
\end{gather}
The equivalence of this regularisation to ordinary QCD,
established in \cite{Frezzotti:2000nk}, is based on axial
transformations of the quark fields (plus the corresponding spurionic
transformations of the mass parameters $m_\ell$ and $\mu_\ell$), which induce
a rotation of composite operators between the two theories.
In particular, for the operator under study one has
\begin{gather}
\label{rotation}
\left(\cO_{\rm VV+AA}\right)_{\rm R}^{\rm QCD} =
\cos(\alpha)\left(\cO_{\rm VV+AA}\right)_{\rm R}^{\rm tmQCD,\alpha} -
i\sin(\alpha)\left(\cO_{\rm VA+AV}\right)_{\rm R}^{\rm tmQCD,\alpha}\ ,
\end{gather}
where the terms have to be interpreted as operator insertions in renormalised Green functions in the continuum limit, and a mass-independent renormalisation scheme is assumed.
Following the notation of \cite{Frezzotti:2000nk}, the twist
angle $\alpha$ depends upon the renormalised mass parameters through the relation $\tan(\alpha)=\mu_{\ell;{\rm R}}/m_{\ell;{\rm R}}$, and (\ref{rotation}) is an identity
holding at each value of $\alpha$. In particular, at $\alpha=\pi/2$,
which is known as the {\it fully twisted} case, (\ref{rotation}) simplifies
to 
\begin{gather}
\label{rotation2}
\left(\cO_{\rm VV+AA}\right)_{\rm R}^{\rm QCD} =
-i\left(\cO_{\rm VA+AV}\right)_{\rm R}^{\rm tmQCD,\pi/2}\ .
\end{gather}
In this way, $\cO_{\rm VV+AA}$ in standard QCD is mapped onto its
counterpart $\cO_{VA+AV}$ in tmQCD.
Using the mass independence of the renormalisation scheme, we will show in the next section that  
$\cO_{\rm VA+AV}$ renormalises multiplicatively in the static
approximation, which represents the main advantage of using the above
mapping. In this sense, the proposed approach represents an extension to
the static case of the tmQCD framework used to determine the $B_K$ parameter~\cite{Palombi:2005zd,Dimopoulos:2004xc}.

\section{Renormalisation pattern}

We now concentrate on the renormalisation properties of 
heavy-light four-quark operators, with the aim of proving that $\cO_{\rm
  VA+AV}$ renormalises multiplicatively. Unfortunately, for brevity's
sake, we skip algebraic details \cite{inpreparation}. We start by
considering generic four-quark operators 
\begin{gather}
O^\pm_{\Gamma_1\Gamma_2} =
    \dfrac{1}{2}\left[(\bar\psi_{h}\Gamma_1\psi_1)(\bar\psi_{\bar
    h}\Gamma_2\psi_2) \pm (\bar\psi_{h}\Gamma_1\psi_2)(\bar\psi_{\bar
    h}\Gamma_2\psi_1)\right],
\end{gather}
where $\Gamma_{1,2}$ represent Dirac matrices. In principle, operators
corresponding to different Dirac structures could mix among them under
renormalisation, thus giving rise to a matrix renormalisation pattern;
consequently a complete basis of such operators must be considered,
such as   
\begin{alignat}{3}
\label{basis}
{\rm parity-even:~~} Q_1^\pm &= O^\pm_{\rm VV+AA}\ ,
\qquad {\rm parity-odd:~~} & {\cQ}_1^\pm &= O^\pm_{\rm VA+AV}\ , \nonumber \\
 Q_2^\pm &= O^\pm_{\rm SS+PP}\ , & {\cQ}_2^\pm &= O^\pm_{\rm SP+PS}\ , \nonumber \\
 Q_3^\pm &= O^\pm_{\rm VV-AA}\ , & {\cQ}_3^\pm &= O^\pm_{\rm VA-AV}\ , \nonumber \\
 Q_4^\pm &= O^\pm_{\rm SS-PP}\ , & {\cQ}_4^\pm &= O^\pm_{\rm SP-PS}\ .
\end{alignat}
The renormalisation matrix $\dZ$, whose size is in principle
$8\times 8$
(mixing between $+$ and $-$ operators is trivially excluded), can be constrained through symmetry
arguments. Given a symmetry of the theory, and the matrix $\Phi$ that
implements a symmetry transformation at the level of the operator basis, it is
sufficient to require that $\dZ$ is invariant under a $\Phi$-rotation
\cite{Becirevic:2003hd}, i.e. 
\begin{gather}
\dZ = \Phi \dZ \Phi^{-1}\ .
\end{gather}
The symmetries we use are:
\begin{itemize}
\item{{\bf Parity}. It prevents the mixing among 
operators with opposite parity. After implementing it, the
renormalisation matrix $\dZ$ is reduced to a block-diagonal form,
where two $4\times 4$ diagonal blocks describe the mixing of the
parity-even and parity-odd operators among themselves.}
\item{{\bf Chiral simmetry}. It is used {\it \`a la}
  \cite{Donini:1999sf}: were chirality respected by the regulator, 
  there  would be no chance of mixing among different chirality
  sectors. The mixing due to the Wilson chirality breaking in the
  parity-odd sector can be represented according to the form
\begin{gather}
\left(\begin{array}{c}
	  \cQ_1^\pm \\
	  \cQ_2^\pm \\
	  \cQ_3^\pm \\
	  \cQ_4^\pm
	\end{array}\right)_{\rm R}
	=
	\left(\begin{array}{cccc}
	  \cZ_{11}^\pm & \cZ_{12}^\pm & 0          & 0          \\
	  \cZ_{21}^\pm & \cZ_{22}^\pm & 0          & 0          \\
	  0          & 0          & \cZ_{33}^\pm & \cZ_{34}^\pm \\
	  0          & 0          & \cZ_{43}^\pm & \cZ_{44}^\pm
	\end{array}\right)
	{\left[ \mathds{1} + 
	\left(\begin{array}{cccc}
	  0               & 0                & \Delta_{13}^\pm & \Delta_{14}^\pm \\
	  0               & 0                & \Delta_{23}^\pm & \Delta_{24}^\pm \\
	  \Delta_{31}^\pm & \Delta_{32}^\pm  & 0               & 0 \\
	  \Delta_{41}^\pm & \Delta_{42}^\pm  & 0               & 0
	\end{array} \right)
	\right]}
	\left(\begin{array}{c}
	  \cQ_1^\pm \\
	  \cQ_2^\pm \\
	  \cQ_3^\pm \\
	  \cQ_4^\pm
	\end{array}\right),
\end{gather}
where the coefficients $\cZ_{ij}$ are scale dependent, while the
$\Delta_{ij}$'s are not. }
\item{{\bf Heavy quark spin symmetry and H(3) spatial rotations}. We  
  then consider two finite spin rotations of the heavy fields,
  plus two lattice spatial rotations of both heavy and light 
  fields 
\begin{align}
& \mbox{heavy quark spin rotations:} & \bar\psi_{h} \to
\bar\psi_{h}\gamma_2\gamma_3, \quad  \bar\psi_{\bar h} \to
\bar\psi_{\bar h}\gamma_2\gamma_3, \nonumber \\
& & \bar\psi_{h} \to
\bar\psi_{h}\gamma_3\gamma_1, \quad  \bar\psi_{\bar h} \to
\bar\psi_{\bar h}\gamma_3\gamma_1, \nonumber \\
& \mbox{lattice spatial rotations:} & \cR(\ \hat 1 \to \hat 2\ ) \ \mbox{rotates the $\hat 1$ axis
      onto the $\hat 2$ axis}, \nonumber \\
& & \cR(\ \hat 1 \to \hat 3\ ) \ \mbox{rotates the $\hat 1$ axis
      onto the $\hat 3$ axis.} 
\end{align}
After a change of basis and some tedious algebra, the parity violating
block reduces to 
\begin{gather}
\label{hqss}
\left(\begin{array}{c}
	  \hskip -0.15cm \cQ_1^\pm \\
	  \hskip -0.15cm \cQ_1^\pm + 4 \hat \cQ_2^\pm \\
	  \hskip -0.15cm \cQ_3^\pm + 2 \hat \cQ_4^\pm \\
	  \hskip -0.15cm \cQ_3^\pm - 2 \hat \cQ_4^\pm
	\end{array}\hskip -0.15cm\right)_{\rm R}
	=
	\left(\begin{array}{cccc}
	  \hskip -0.15cm \cZ_{1}^\pm & \hskip -0.15cm 0           & \hskip -0.15cm 0           & \hskip -0.15cm 0          \\
	  \hskip -0.15cm 0           & \hskip -0.15cm \cZ_{2}^\pm & \hskip -0.15cm 0           & \hskip -0.15cm 0          \\
	  \hskip -0.15cm 0           & \hskip -0.15cm 0           & \hskip -0.15cm \cZ_{3}^\pm & \hskip -0.15cm 0 \\
	  \hskip -0.15cm 0           & \hskip -0.15cm 0           & \hskip -0.15cm 0           & \hskip -0.15cm \cZ_{4}^\pm
	\end{array}\hskip -0.15cm \right)
	{\left[ \mathds{1} + 
	\left(\begin{array}{cccc}
	  \hskip -0.15cm 0              & 0               & \Delta_{1}^\pm & 0 \\
	  \hskip -0.15cm 0              & 0               & 0              & \Delta_{2}^\pm \\
	  \hskip -0.15cm \Delta_{3}^\pm & 0               & 0              & 0 \\
	  \hskip -0.15cm 0              & \Delta_{4}^\pm  & 0              & 0
	\end{array}\hskip -0.15cm \right)
	\right]}
	\left(\begin{array}{c}
	  \hskip -0.15cm \cQ_1^\pm \\
	  \hskip -0.15cm \cQ_1^\pm + 4 \cQ_2^\pm \\
	  \hskip -0.15cm \cQ_3^\pm + 2 \cQ_4^\pm \\
	  \hskip -0.15cm \cQ_3^\pm - 2 \cQ_4^\pm
	\end{array}\hskip -0.15cm\right) .
\end{gather}
}
\item{{\bf Time reversal}. We finally consider a time reversal
  transformation of the quark fields:
\begin{gather}
 \psi_h(x) \to \gamma_0\gamma_5\psi_{\bar h}(x^\tau),\quad \psi_{\bar h}(x) \to \gamma_0\gamma_5\psi_h(x^\tau), \quad
 \psi_k(x)\to \gamma_0\gamma_5\psi_k(x^\tau),\quad k=1,2.
\end{gather}
It further constrains the parity-odd block by forcing the residual
$\Delta_i$ coefficients in (3.6) to vanish.
Purely multiplicative renormalisation of $\cQ_{\rm VA+AV}^\pm$ follows therefrom. 
}
\end{itemize}

\section{Renormalisation in Schr\"odinger Functional schemes}

We use the Schr\"odinger Functional (SF) to define a family of finite
volume renormalisation schemes, in view of a non-perturbative
study of the running of the $\cO_{VA+AV}$ operator. Our approach
closely follows here refs. \cite{Palombi:2005zd}, to
which the reader is referred for unexplained notation. We first
introduce bilinear boundary sources at $x_0=0,T$ (being $T$ the time
extension of the SF),
\begin{align}
\label{source}
\cS_{\fla_1\fla_2}[\Gamma] &=
a^6\sum_{\vx,\vy}\bar{\zeta}_{\fla_1}(\vx)\Gamma\zeta_{\fla_2}(\vy) \
,\qquad \cS'_{\fla_1\fla_2}[\Gamma] = a^6\sum_{\vx,\vy}\bar{\zeta}'_{\fla_1}(\vx)\Gamma\zeta'_{\fla_2}(\vy) \ ,
\end{align}
where $\Gamma$ is a Dirac matrix and the flavour indices $\fla_{1,2}$
can assume either relativistic or static values. Then, we define a set of
SF correlators in order to probe the operators $\cQ^\pm_1$, 
\begin{align}
& F_1^\pm(x_0) = \frac{a^3}{L^3}\sum_{\bf x} \langle
\cS'_{3\bar
  h}[\Gamma_3]{\cQ}_1^\pm(x)\cS_{1h}[\Gamma_1]\cS_{23}[\Gamma_2]\rangle\ , \nonumber \\
& f_1^{\fla_1\fla_2} = -\frac{1}{L^6}\langle \cS'_{\fla_1\fla_2}[\gamma_5]\cS_{\fla_2\fla_1}[\gamma_5]\rangle\ ,
\qquad k_1^{\fla_1\fla_2} = -\frac{1}{3L^6}\sum_{k=1}^3\langle
\cS'_{\fla_1\fla_2}[\gamma_k]\cS_{\fla_2\fla_1}[\gamma_k]\rangle\ .
\end{align}
The triple $[\Gamma_1,\Gamma_2,\Gamma_3]$ has to be chosen such that
$F_1^\pm$ is non-zero. The boundary correlators $f_1$ and $k_1$, which
can have either {\it light-light} or {\it heavy-light} flavour structure,
are needed in order to cancel the renormalisation of the boundary
sources in $F_1^\pm$. In practice, we consider ratios
of the form
\begin{gather}
\label{ratio}
h_1^\pm (x_0) = \frac{F_1^\pm(x_0)}{f_1^{hl}\ [f_1^{ll}]^{\alpha}\
  [k_1^{ll}]^{1/2-\alpha}\ }
\end{gather}
and then impose in the chiral limit the renormalisation condition
\begin{gather}
\cZ_1^\pm h_1^\pm(T/2) = h_1^\pm(T/2)|_{g_0=0}.
\end{gather}
Of course, the renormalisation factor $\cZ_1^\pm$ depends upon all
calculational details, e.g. the light quark action (Wilson
with(out) a clover term), the static action (Eichten-Hill, or its ALPHA
variants), the choice of the Dirac structures $[\Gamma_1,\Gamma_2,\Gamma_3]$, the value of
the $\theta$-angle of the SF and the value of the parameter $\alpha$
introduced in (\ref{ratio}). This richness of degrees of freedom 
can be exploited in order to identify some optimal renormalisation
schemes, according to the general requirements of maximisation of the
nonperturbative signal/noise ratio, slowing down of the operator
running and minimisation of lattice artefacts. 

\section{NLO anomalous dimension of $\cQ_1^+$ from perturbative
  matching at one-loop order} 

In order to gain information about the running and its lattice
artefacts, we have performed a one-loop perturbative calculation of
the renormalisation factor $\cZ^+_{1}$ in some of the SF schemes
discussed above. Such a calculation allows us to determine the NLO
anomalous dimension of $\cQ_1^+$ via a perturbative matching 
to some reference scheme in which the NLO anomalous dimension is
already known.
The matching procedure has been illustrated and applied several times
in the literature \cite{Sint:1998iq,Palombi:2005zd,Guagnelli:2003hw},
and it will not be reviewed here. The reference scheme was chosen to
be DRED, where the NLO anomalous dimension of $\cQ_1^+$ and its 
perturbative matching to the so-called {\it lat}-scheme have been
computed in \cite{Gimenez:1998mw}. The perturbative
expansion of $\cZ_1^+$ reads
\begin{gather}
\cZ_1^+(g_0^2,a/L) = 1 + \sum_{k=1}^\infty g_0^{2k}\cZ_1^{+(k)} = 1 + g_0^2\left[r_0^+ +\gamma_0^+\ln\left(\frac{a}{L}\right) +
  O\left(\frac{a}{L}\right)\right] + O(g_0^4)
\end{gather}
where $\gamma_0^+=-1/(2\pi^2)$ is the universal anomalous dimension
of the operator $\cQ_1^+$, and $r_0^+$ is the one-loop scheme-dependent
finite part, peculiar to the SF and the defining choices listed at the
end of the previous section. The running of the operator is described
by the step scaling function (ssf),
\begin{gather}
\sigma_1^+(u) =
\lim_{a\to0}\frac{\cZ_1^+(g_0,a/2L)}{\cZ_1^+(g_0,a/L)}\biggr|_{\bar
  g^2(L)=u} = 1 + \sum_{k=1}^\infty g_0^{2k}\sigma_1^{+(k)}
\end{gather}
As an example of the running, the ssf of $\cQ_1^+$ at NLO
and $N_{\rm f}=0$ is reported as a function of the renormalised
coupling on the left side of Figure~1. The plot refers to the choice 
$[\Gamma_1,\Gamma_2,\Gamma_3]=[\gamma_5,\gamma_5,\gamma_5]$. The 
straight line represents the universal LO running, and the bands describe
the dependence of the NLO anomalous dimension upon the choice of
$\alpha$, when the latter ranges in the interval $[0,1/2]$. On the right
side of Figure~1 we report a comparison of the lattice artefacts
$\delta^+_1(a/L)$ on the ssf, defined as in \cite{Sint:1998iq}, between the
static-light case and the light-light one (data from
\cite{Palombi:2005zd}). The comparison refers to the schemes where
$[\Gamma_1,\Gamma_2,\Gamma_3]=[\gamma_5,\gamma_5,\gamma_5]$,
$\theta=0.5$ and $\alpha=0$. The light
quarks are discretised according to the unimproved (W) or the $c_{\rm
 sw}$-improved (SW) Wilson action, while the static quarks are
discretised according to the Eichten-Hill (EH) action. Although the
static-light schemes cannot be directly compared to the relativistic
ones (where the normalisation of the four-quark correlator is
always performed using only the relativistic correlators),
the plot shows that the introduction of static quarks does not imply
a significant increment of the lattice artefacts in perturbation theory.

\section{Acknowledgments}

We thank D.~Be\'cirevi\'c, M.~Della~Morte, R.~Sommer and especially
J.~Reyes for useful discussions. F.P. acknowledges the Conference
organisers and the Humboldt-Foundation for financial support. 

\begin{figure}
\begin{center}
\begin{minipage}{.45\linewidth}
\psfrag{xlabel}[c][bc][1][0]{$T/a$}
\psfrag{ylabel}[c][c][1][0]{$\log_{10}r$}
\begin{center}
\epsfig{scale=.29,file=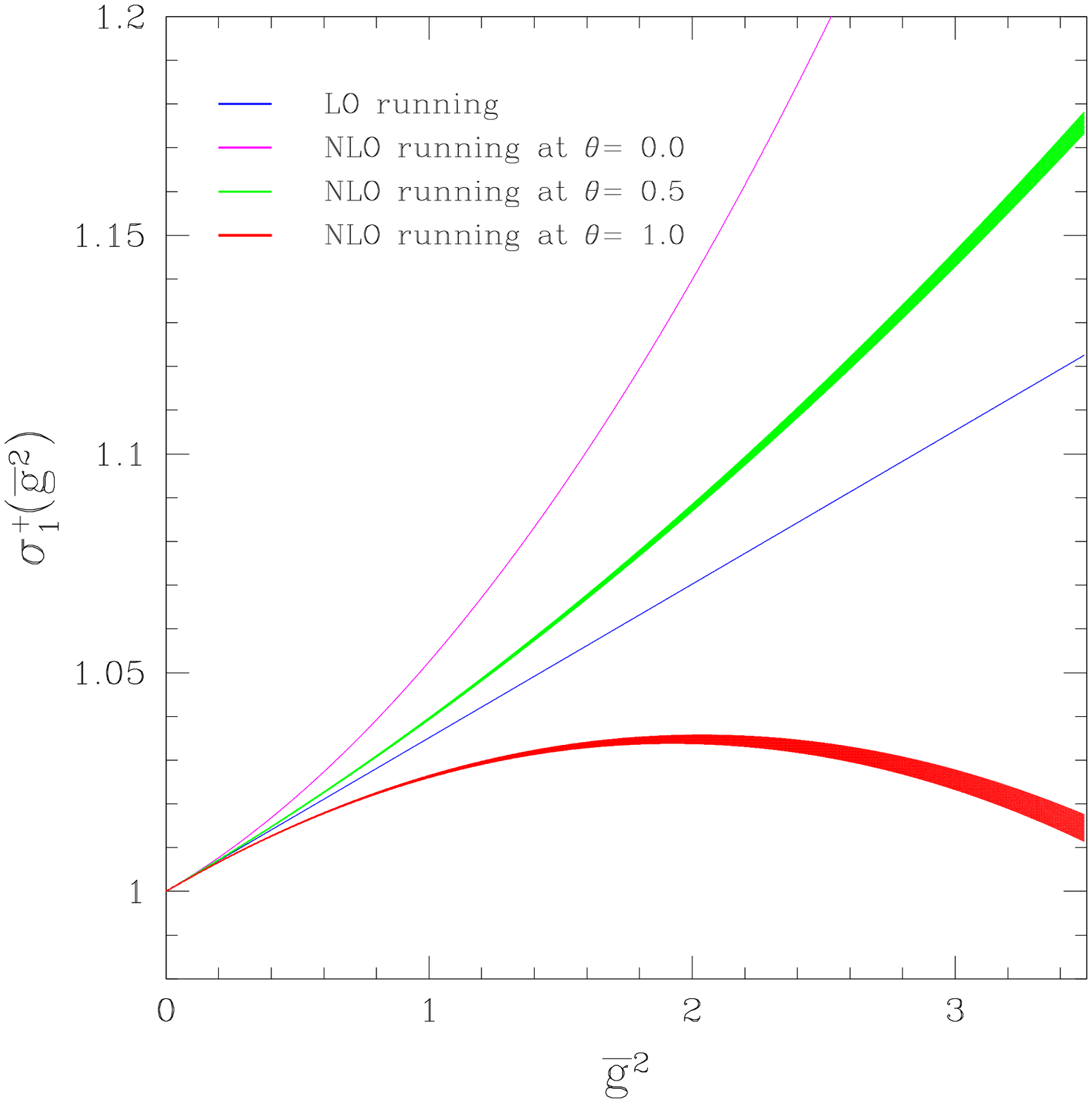}
\end{center}
\end{minipage}
\begin{minipage}{.45\linewidth}
\psfrag{xlabel}[c][bc][1][0]{$T/a$}
\psfrag{ylabel}[c][c][1][0]{$\log_{10}r$}
\begin{center}
\epsfig{scale=.29,file=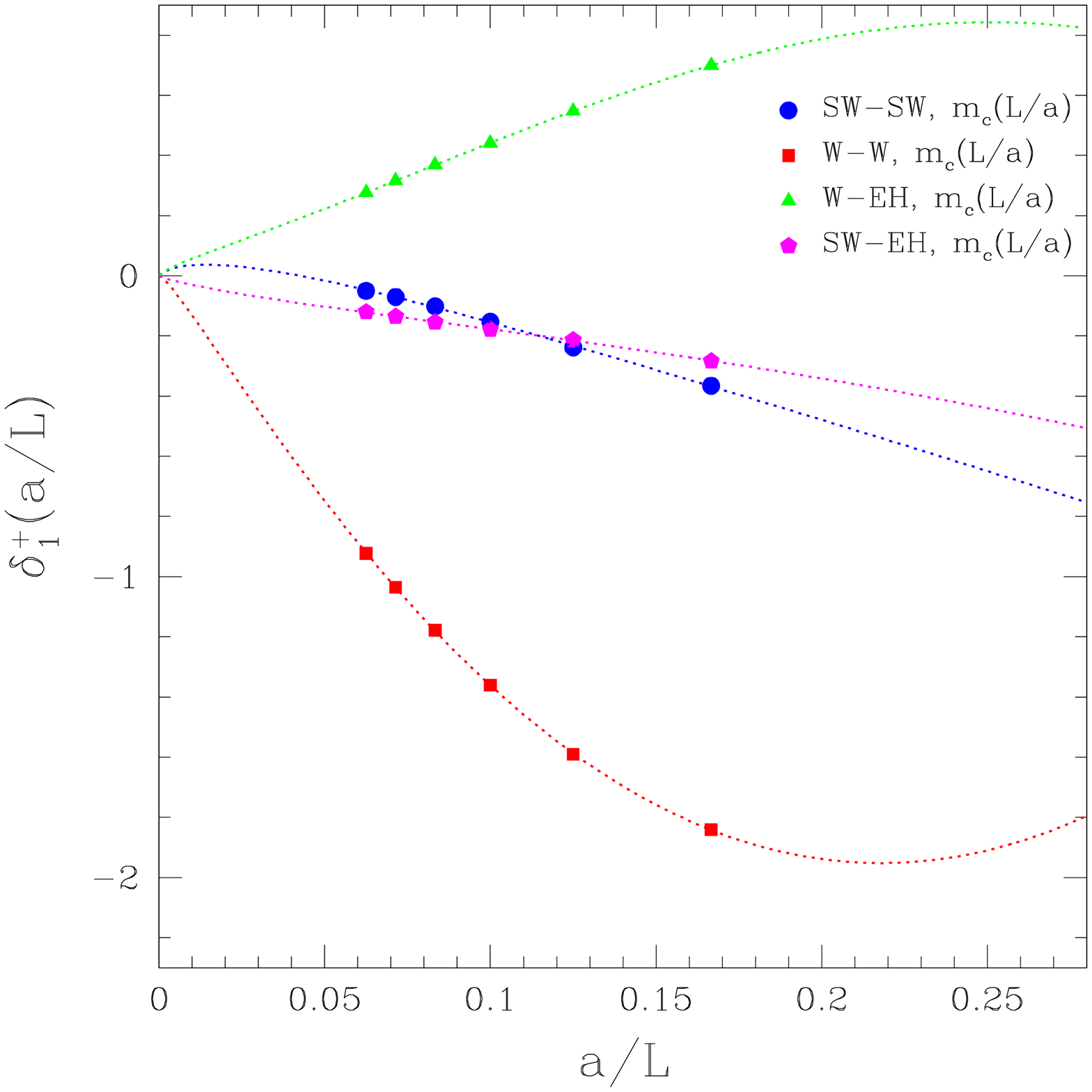}
\end{center}
\end{minipage}
\caption{On the left side the step scaling function of $\cQ_1^+$ at NLO
  and $N_{\rm f}=0$ is reported vs. the renormalised coupling in the
  SF scheme. On the right side we compare the lattice artefacts of the
  step scaling function between the full relativistic case and the
  static-light case. Both plots are preliminary.}
\end{center}
\vspace{-.4cm}
\end{figure}

\end{document}